\documentclass[twocolumn]{aastex631}

\usepackage{amsmath}

\begin{document}

\title{JWST measurements of $^{13}$C, $^{18}$O and $^{17}$O in the atmosphere of super-Jupiter VHS~1256~b}

\correspondingauthor{Siddharth Gandhi}
\email{Siddharth.Gandhi@warwick.ac.uk}

\author[0000-0001-9552-3709]{Siddharth Gandhi}
\affiliation{Leiden Observatory, Leiden University, Postbus 9513, 2300 RA Leiden, The Netherlands}
\affiliation{Department of Physics, University of Warwick, Coventry CV4 7AL, UK}
\affiliation{Centre for Exoplanets and Habitability, University of Warwick, Gibbet Hill Road, Coventry CV4 7AL, UK}

\author[0000-0003-4760-6168]{Sam de Regt}
\affiliation{Leiden Observatory, Leiden University, Postbus 9513, 2300 RA Leiden, The Netherlands}

\author{Ignas Snellen}
\affiliation{Leiden Observatory, Leiden University, Postbus 9513, 2300 RA Leiden, The Netherlands}

\author{Yapeng Zhang}
\affiliation{Leiden Observatory, Leiden University, Postbus 9513, 2300 RA Leiden, The Netherlands}

\author{Benson Rugers}
\affiliation{Leiden Observatory, Leiden University, Postbus 9513, 2300 RA Leiden, The Netherlands}

\author{Niels van Leur}
\affiliation{Leiden Observatory, Leiden University, Postbus 9513, 2300 RA Leiden, The Netherlands}

\author{Quincy Bosschaart}
\affiliation{Leiden Observatory, Leiden University, Postbus 9513, 2300 RA Leiden, The Netherlands}


\begin{abstract}
Isotope ratios have recently been measured in the atmospheres of directly-imaged and transiting exoplanets from ground-based observations. The arrival of JWST allows us to characterise exoplanetary atmospheres in further detail and opens up wavelengths inaccessible from the ground. In this work we constrain the carbon and oxygen isotopes $^{13}$C, $^{18}$O and $^{17}$O from CO in the atmosphere of the directly-imaged companion VHS~1256~b through retrievals of the $\sim$4.1-5.3~$\mu$m NIRSpec G395H/F290LP observations from the early release science programme (ERS 1386). We detect and constrain $^{13}$C$^{16}$O, $^{12}$C$^{18}$O and $^{12}$C$^{17}$O at 32, 16 and 10$\sigma$ confidence respectively, thanks to the very high signal-to-noise observations. We find the ratio of abundances are more precisely constrained than their absolute values, with $\mathrm{^{12}C/^{13}C=62^{+2}_{-2}}$, in between previous measurements for companions ($\sim$30) and isolated brown dwarfs ($\sim$100). The oxygen isotope ratios are $\mathrm{^{16}O/^{18}O =425^{+33}_{-28}}$ and $\mathrm{^{16}O/^{17}O=1010^{+120}_{-100}}$. All of the ratios are lower than the local inter-stellar medium and Solar System, suggesting that abundances of the more minor isotopes are enhanced compared to the primary. This could be driven by isotope fractionation in protoplanetary disks, which can potentially alter the carbon and oxygen ratios through isotope selective photodissociation, gas/ice partitioning and isotopic exchange reactions. In addition to CO, we constrain $^{1}$H$_2$$^{16}$O and $^{12}$C$^{16}$O$_2$ (the primary isotopologues of both species), but find only upper limits on $^{12}$C$^1$H$_4$ and $^{14}$N$^{1}$H$_3$. This work highlights the power of JWST to constrain isotopes in exoplanet atmospheres, with great promise in determining formation histories in the future.
\end{abstract}



\section{Introduction} \label{sec:intro}

Isotope ratios in exoplanetary atmospheres are a new frontier in high-resolution spectroscopy. First proposed by \citet{molliere2019} and \citet{morley2019}, atmospheric isotopes have been measured for a range of directly-imaged companions and brown dwarf atmospheres from ground-based observations \citep{zhang2021_bd, zhang2021_nature}, as well as in one close-in transiting exoplanet \citep{line2021}. These isotopic constraints have hinted that $^{12}$C/$^{13}$C ratios vary between targets, and recent work has also tentatively constrained the oxygen isotope ratio $^{16}$O/$^{18}$O in a brown dwarf \citep{zhang2022}, suggesting a super-solar value. This opens up a key dimension towards tracing planet formation and migration history through isotope fractionation processes.

The arrival of JWST represents a significant paradigm change for exoplanet science, with a range of instruments capable of characterising atmospheres at high signal-to-noise over a wide wavelength range across the infrared. The NIRSpec spectrograph in particular is able to observe at wavelengths difficult to observe from the ground due to heavy telluric absorption and reaches spectral resolutions R$\sim$2,700. Results from the Early Release Science programmes for both transiting and directly-imaged planets has shown key species such as H$_2$O, CO$_2$, CO, SO$_2$ and CH$_4$ in exoplanet atmospheres \citep[e.g.,][]{jwst2023, constantinou2023, anderson2023, miles2023, grant2023}. This is set to continue in the next decade and will allow us to obtain some of the most precise constraints on trace species for both close-in transiting and more widely separated directly-imaged exoplanets. 

Atmospheric characterisation of directly-imaged exoplanets has grown extensively in recent years thanks to high signal-to-noise observations \citep[e.g.][]{chilcote2017, samland2017}. Retrievals of such targets are often more complex than more commonly studied transiting exoplanets, requiring differing parametrisations of the temperature profile \citep[e.g.,][]{line2015, piette2020}, and additional free parameters for the radius and surface gravity given that the masses and radii are often not well known. A number of targets have been characterised with direct imaging \citep[e.g.,][]{janson2013, todorov2016}, most notably the HR~8799 system \citep{lee2013, lavie2017, molliere2020}. Direct imaging is also of great importance for the future of atmospheric characterisation, as it is the most likely means by which we will be able to characterise atmospheres of Earth-like planets around Sun-like stars with future generation facilities such as the Habitable Worlds Observatory \citep{luvoir2019, gaudi2020, decadalsurvey2021} and Large Interferometer For Exoplanets \citep{quanz2022}.

\begin{figure*}
\centering
	\includegraphics[width=\textwidth,trim={0cm 0cm 0cm 0},clip]{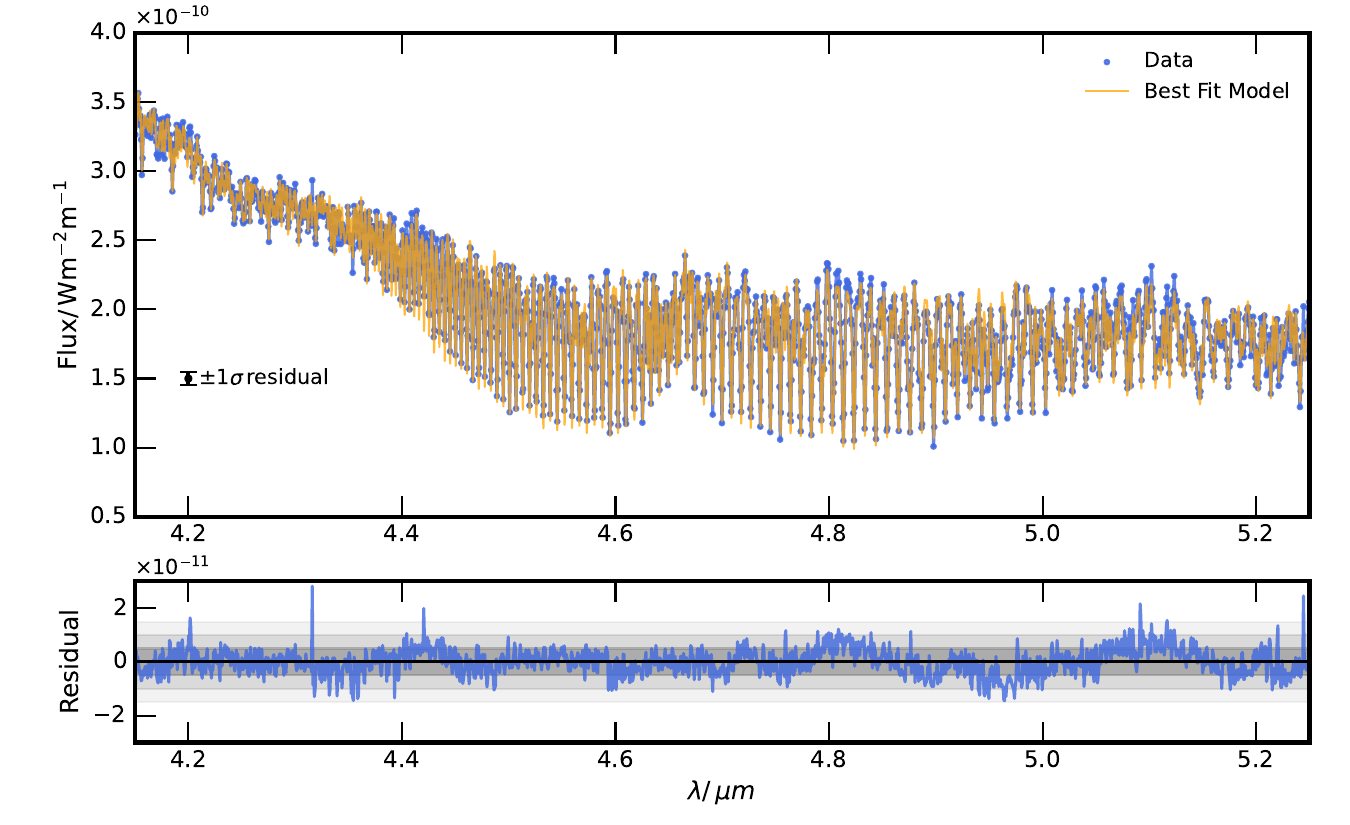}
    \caption{Top Panel: Best fitting model from the retrieval, along with the data points of the NIRSpec G395H/F290LP detector 2 (NRS2) observations of VHS~1256~b. We also show the 1$\sigma$ range of the residuals of the data and model. Bottom Panel: Residuals of the best fit model and the observations, with the 1, 2 and 3$\sigma$ ranges shown in grey.}     
\label{fig:bestfit}
\end{figure*}

In this work we retrieve the spectrum of VHS~J125601.92–125723.9~b (hereafter VHS~1256~b), a directly-imaged companion, also called a super-Jupiter, that straddles the brown dwarf/planet boundary \citep{gauza2015, dupuy2023} at the edge of the L/T transition. We use data from the JWST early release science programme (ERS 1386, PI Hinkley) \citep{miles2023}, performing spectral retrievals of the G395H/F290LP NRS 2 observations which cover the 4.1-5.3~$\mu$m wavelength range. We explore the isotopic ratios of CO by separately retrieving the chemical abundances of $^{12}$C$^{16}$O, $^{13}$C$^{16}$O, $^{12}$C$^{18}$O and $^{12}$C$^{17}$O, in addition to H$_2$O, CO$_2$, CH$_4$ and NH$_3$. In particular, $^{12}$C/$^{13}$C is an important quantity given that it has shown some variation between exoplanets and brown dwarfs \citep{zhang2021_bd} and shows differences between the Solar System and the local inter-stellar medium \citep[ISM;][]{wilson1999, milam2005}.

The next section discusses the retrieval setup, followed by the results, and finally we present the discussion and conclusions of our work.

\section{Atmospheric Retrieval}

We perform our atmospheric retrieval with HyDRA \citep{gandhi2018, gandhi2023}, modified to allow for the characterisation of non-irradiated exoplanets, as discussed below. We use the JWST NIRSpec IFU G395H/F290LP dataset of VHS~1256~b \citep{miles2023}, covering the $\sim$4.1-5.3~$\mu$m wavelength range of NRS2. This is the wavelength range where CO and all of its isotopologues have prominent opacity. In addition, CO has well separated spectral lines which are resolvable at the R$\sim$2,700 spectral resolution of the observations, and the $^{12}$C$^{16}$O absorption lines are clearly visible in the observed spectrum shown in Figure~\ref{fig:bestfit}. Henceforth, unless otherwise stated, we refer to $^{12}$C and $^{16}$O as C and O respectively, as these are the primary isotopes of each species.

\begin{table}
\begin{center}
\caption{Temperature and pressure grid for the cross sections for each species.}\label{table:cross_sec_grid}
\begin{tabular}{ c| c c c c c c} 
\hline
 \textbf{T(K)} & 300&400&500&600&700&800 \\
 &900&1000&1200&1400&1600&1800 \\ 
 &2000&2500&3000&3500&4000 \\
 \hline
 \textbf{P(bar)} & $10^{-5}$ & $10^{-4}$ & $10^{-3}$ & $10^{-2}$ & $10^{-1}$&1\\
 &10&100 \\
 \hline
\end{tabular}
\end{center}
\end{table}

\subsection{Molecular cross sections}
We retrieve the chemical abundances of CO (main isotopologue), $^{13}$CO, C$^{18}$O, C$^{17}$O, H$_2$O, CO$_2$, CH$_4$ and NH$_3$. The CO isotopologues all have differing cross sections as their line positions are shifted given the differing masses of the carbon and oxygen atoms. Hence each isotopologue is independently distinguishable in the spectrum given sufficient spectral resolution and signal-to-noise. We use the main isotopologue only for species other than CO, with the exception of CH$_4$, where we include the opacity for all isotopologues weighted by their terrestrial abundances. We use line lists from ExoMol \citep{tennyson2016} for H$_2$O \citep{polyansky2018} and NH$_3$ \citep{coles2019}, HITEMP for the CO isotopologues \citep{rothman2010, li2015} and CH$_4$ \citep{hargreaves2020}, and Ames for CO$_2$ \citep{huang2013, huang2017}. We calculate the molecular cross section of each species on a grid of pressure and temperature (see Table~\ref{table:cross_sec_grid}), with each line broadened by applying a Voigt profile given the H$_2$/He broadening coefficients. Further details of the cross section calculations can be found in \citet{gandhi2020_cs}. We set the vertically constant volume mixing ratios of each species as a free parameter in the retrieval, resulting in eight free parameters for the chemical composition of the atmosphere. Vertically constant abundances are justified given that the abundances of H$_2$O and CO are not expected to vary significantly with pressure, and because observations from other similar targets have shown that vertically-varying chemistry such as with chemical equilibrium models do not necessarily provide as good a fit as those with free chemistry \citep{deregt2023}. In addition to the molecular opacity, we also include absorption from H$_2$-H$_2$ and H$_2$-He collisionally induced absorption \citep{richard2012}.

\subsection{P-T profile}\label{sec:PT}

We parametrise the P-T profile according to the method in \citet{deregt2023}, with 9 P-T knots, spaced such that grid points are more frequent in the photosphere region between 10-0.1~bar (see Figure~\ref{fig:PT}). This ensures that there are a sufficient number of knots to cover the full atmosphere but keeps their overall number to a minimum in the regions where the spectrum is less sensitive. To prevent oscillatory behaviour, the general P-splines formalism of \citet{li2022} is employed to compute a log-likelihood penalty as
\begin{equation}
    \ln\mathcal{L}_\mathrm{penalty} = -\frac{\mathrm{PEN}^{(3)}_\mathrm{gps}}{2\gamma}  - \frac{1}{2}\ln(2\pi\gamma). 
\end{equation}
Here, $\mathrm{PEN}^{(3)}_\mathrm{gps}$ is the third-order general difference penalty which increases for larger temperature variations at the P-T knots. Additionally, temperature variations are penalised more severely for smaller knot separations in log pressure. Similar to \citet{line2015}, the factor $\gamma$ is a retrieved parameter that scales the contribution of the penalty to the overall log-likelihood, thereby allowing for more oscillations if the data warrants it. In total, 10 free parameters are utilised for the P-T profile. The temperature profile used to generate the planetary spectrum incorporates 50 layers between 100-10$^{-6}$~bar evenly spaced in log pressure.

\begin{table}
    \centering
    \begin{tabular}{c|c|c}
& \textbf{Parameter}              & \textbf{Prior Range}\\
\hline
Chemistry   & $\log(\mathrm{H_2O})$  & -12 $\rightarrow$ -1 \\
            & $\log(\mathrm{^{12}CO})$ & -12 $\rightarrow$ -1 \\
            & $\log(\mathrm{^{13}CO})$ & -12 $\rightarrow$ -1 \\
            & $\log(\mathrm{C^{18}O})$ & -12 $\rightarrow$ -1 \\
            & $\log(\mathrm{C^{17}O})$ & -12 $\rightarrow$ -1 \\
            & $\log(\mathrm{CO_2})$ & -12 $\rightarrow$ -1 \\
            & $\log(\mathrm{CH_4})$ & -12 $\rightarrow$ -1 \\
            & $\log(\mathrm{NH_3})$ & -12 $\rightarrow$ -1 \\
\hline
Temp. Profile & $T_\mathrm{100bar}$ / K & 300 $\rightarrow$ 4000 \\
            & $T_\mathrm{10bar}$ / K & 300 $\rightarrow$ 4000 \\
            & $T_\mathrm{3bar}$ / K & 300 $\rightarrow$ 4000 \\
            & $T_\mathrm{1bar}$ / K & 300 $\rightarrow$ 4000 \\
            & $T_\mathrm{0.3bar}$ / K & 300 $\rightarrow$ 4000 \\
            & $T_\mathrm{0.1bar}$ / K & 300 $\rightarrow$ 4000 \\
            & $T_\mathrm{0.01bar}$ / K & 300 $\rightarrow$ 4000 \\
            & $T_\mathrm{10^{-4}bar}$ / K & 300 $\rightarrow$ 4000 \\
            & $T_\mathrm{10^{-6}bar}$ / K & 300 $\rightarrow$ 4000 \\
\hline
P-T penalty & $\log(\gamma)$ & -3 $\rightarrow$ 2 \\
\hline
Planet param. & $\mathrm{R_p/R_J}$ & 0.6 $\rightarrow$ 2.5 \\
            & $\log(\mathrm{g/cms^{-2}})$ & 3.5 $\rightarrow$ 5.5 \\
\hline
Clouds/hazes & $\log(\kappa_\mathrm{4.1\mu m}/\mathrm{cm^2g^{-1}})$ & -10 $\rightarrow$ 2 \\ 
            & $\log(\kappa_\mathrm{5.3\mu m}/\mathrm{cm^2g^{-1}})$ & -10 $\rightarrow$ 2 \\ 
            & $\log(P_\mathrm{cl}/\mathrm{bar})$ & -4 $\rightarrow$ 2 \\
            & $\log(\alpha_\mathrm{cl})$ & 0 $\rightarrow$ 20 \\
\hline 
            & $\log(\beta_\mathrm{err})$ & -1 $\rightarrow$ 2 \\
            & $\Delta V_\mathrm{sys}$ / kms$^{-1}$ & -30 $\rightarrow$ 30 \\
            & resolution & 1500 $\rightarrow$ 5000 \\
            & $f_\lambda$ & 0.7 $\rightarrow$ 1.3 \\
    \end{tabular}
    \caption{Parameters and uniform prior ranges for our retrieval of VHS~1256~b.}
    \label{tab:priors}
\end{table}

\subsection{Retrieval Setup}
In addition to the chemical abundances and temperature profile parameters, we retrieve the planetary radius, log gravity, the spectral resolution of the observations, planetary velocity shift, an error scale factor and a wavelength-dependent model scale factor. The error scale factor, $\beta_\mathrm{err}$ allows for a scaling of all of the error bars of the data points. The overall likelihood per model is determined as
\begin{align}
    \log(\mathcal{L}) = &-\frac{1}{2} \sum_i \Big( \frac{d_i - m_i}{\beta_\mathrm{err} \sigma_i} \Big)^2 - \frac{1}{2}\sum_i \ln(2\pi (\beta_\mathrm{err} \sigma_i)^2) \\
    &+ \ln\mathcal{L}_\mathrm{penalty},\nonumber
\end{align}
where $d_i$ and $m_i$ refer to the data and model for each point $i$, and $\sigma_i$ represents the uncertainty of each data point.

The wavelength-dependent model scale factor, $f_\lambda$, scales the model linearly with the wavelength relative to the blue end of the spectrum. The scaled spectrum used in the retrieval is given by
\begin{equation}
    F_\mathrm{scaled}(\lambda) = F_\mathrm{unscaled}(\lambda) \Big( 1 + \frac{\lambda - 4.1~\mathrm{\mu m}}{5.3~\mathrm{\mu m} - 4.1~\mathrm{\mu m}} f_\lambda \Big) .
\end{equation}
This new approach was implemented given the high signal-to-noise of the observations, and allows flexibility for any gradients in the spectrum that cannot be well modelled and/or any systematics in the spectrum. In addition, the spectral resolution of the observations varies with wavelength, and such a formulation allows some flexibility of the model to account for this. For instance, an increasing spectral resolution with wavelength would be compensated by a positive value for $f_\lambda$, as the redder wavelengths are more highly weighted in the convolution.

We also retrieve a non-grey cloud deck through four additional parameters, with cloud opacities set at 4.1 and 5.3~$\mu$m and interpolated at wavelengths in between, as well as a cloud deck pressure, $P_\mathrm{cl}$, and pressure-dependent power-law, $\alpha_\mathrm{cl}$ \citep[see e.g.,][]{molliere2020, deregt2023}. Clouds have been shown to be significant in this target \citep{miles2023}, and our cloud model ensures the cloud deck is sufficiently flexible to account a range of cloud deck pressures as well as any wavelength dependent variation in its opacity across the $\sim$1.2~$\mu$m range of the observations.

Overall, the retrieval therefore includes 28 free parameters, with eight for the volume mixing ratio of each chemical species (including the CO isotopologues) and nine for the P-T profile, with an additional penalty scale factor (see Table~\ref{tab:priors}). Our prior ranges were chosen to allow the retrieval to explore a wide range of potential solutions for the atmosphere, and ensured that none of the parameters converged to the edge of the prior range. Note however that the choice of prior range does impact the overall evidence. For the detection significances, we removed the specific free parameter of the volume mixing ratio of the given species, and ensured that the other parameters in the retrieval were left with the same prior. Each model is generated at a spectral resolution of R=100,000, before being convolved by a Gaussian profile to the spectral resolution of the observations (a free parameter in our retrieval). We perform the retrievals with MultiNest \citep{feroz2008, feroz2009, buchner2014}. We also perform additional retrievals by removing each species at a time for the leave-one-out cross-validation tests (see section~\ref{sec:robustness}) and in order to determine the detection significances of each species.

\section{Results}

\begin{figure}
\centering
	\includegraphics[width=\columnwidth,trim={0cm 0cm 0cm 0},clip]{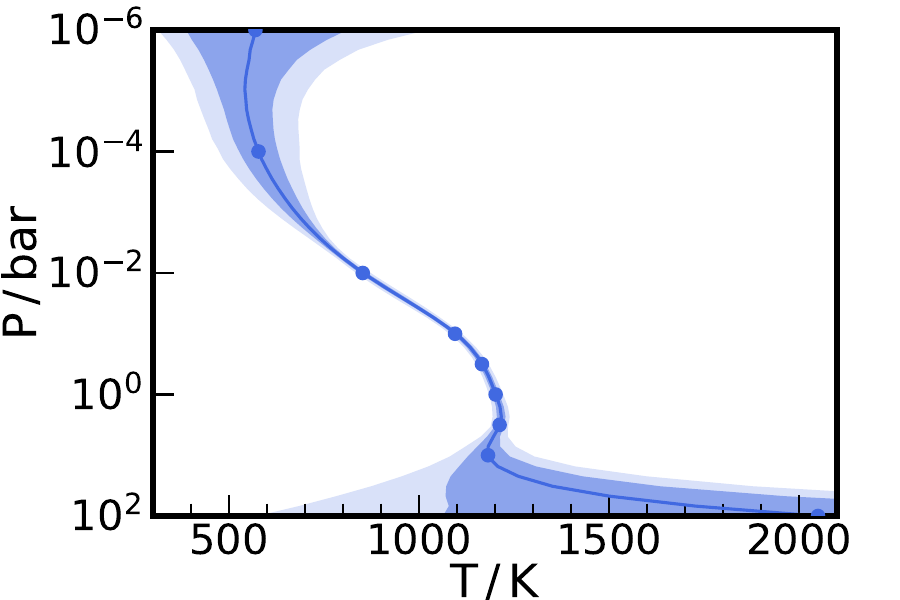}
    \caption{Retrieved pressure-temperature profile form the retrieval of VHS~1256~b. The solid line shows the median values and the dark and light shaded regions indicate the 1 and 2$\sigma$ uncertainties respectively. The markers indicate the P-T knots used in the retrieval (see section~\ref{sec:PT}).}     
\label{fig:PT}
\end{figure}

\begin{figure}
\centering
	\includegraphics[width=\columnwidth,trim={0cm 0cm 0cm 0},clip]{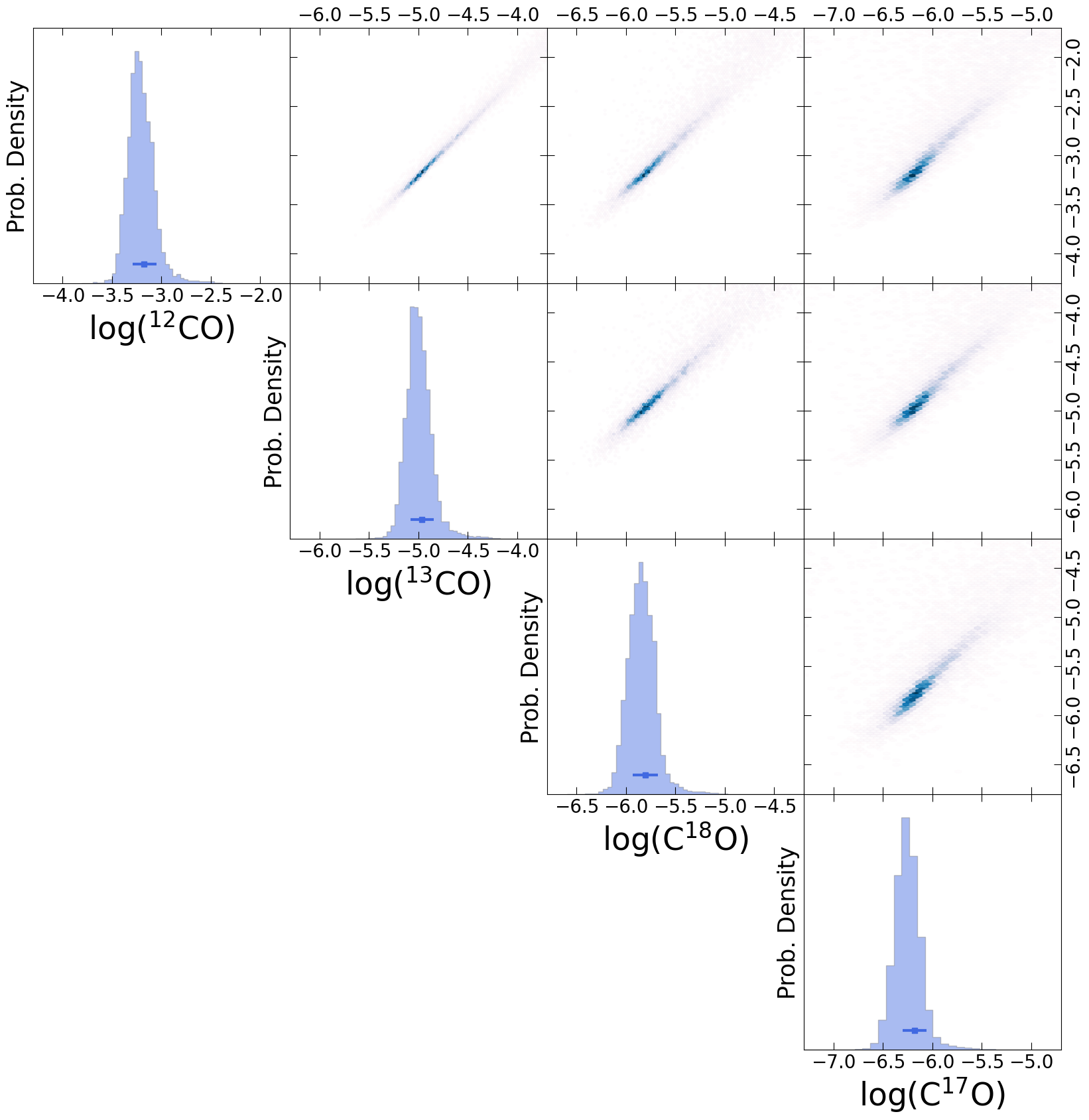}
    \caption{Posterior distributions of the volume mixing ratios of the isotopologues of CO in the retrieval of VHS~1256~b. These show the dependence of one species on the other. Note that C and O refer to the main isotopologues, $^{12}$C and $^{16}$O, respectively.}     
\label{fig:posterior}
\end{figure}

\begin{figure*}
\centering
	\includegraphics[width=\textwidth,trim={0cm 0cm 0cm 0},clip]{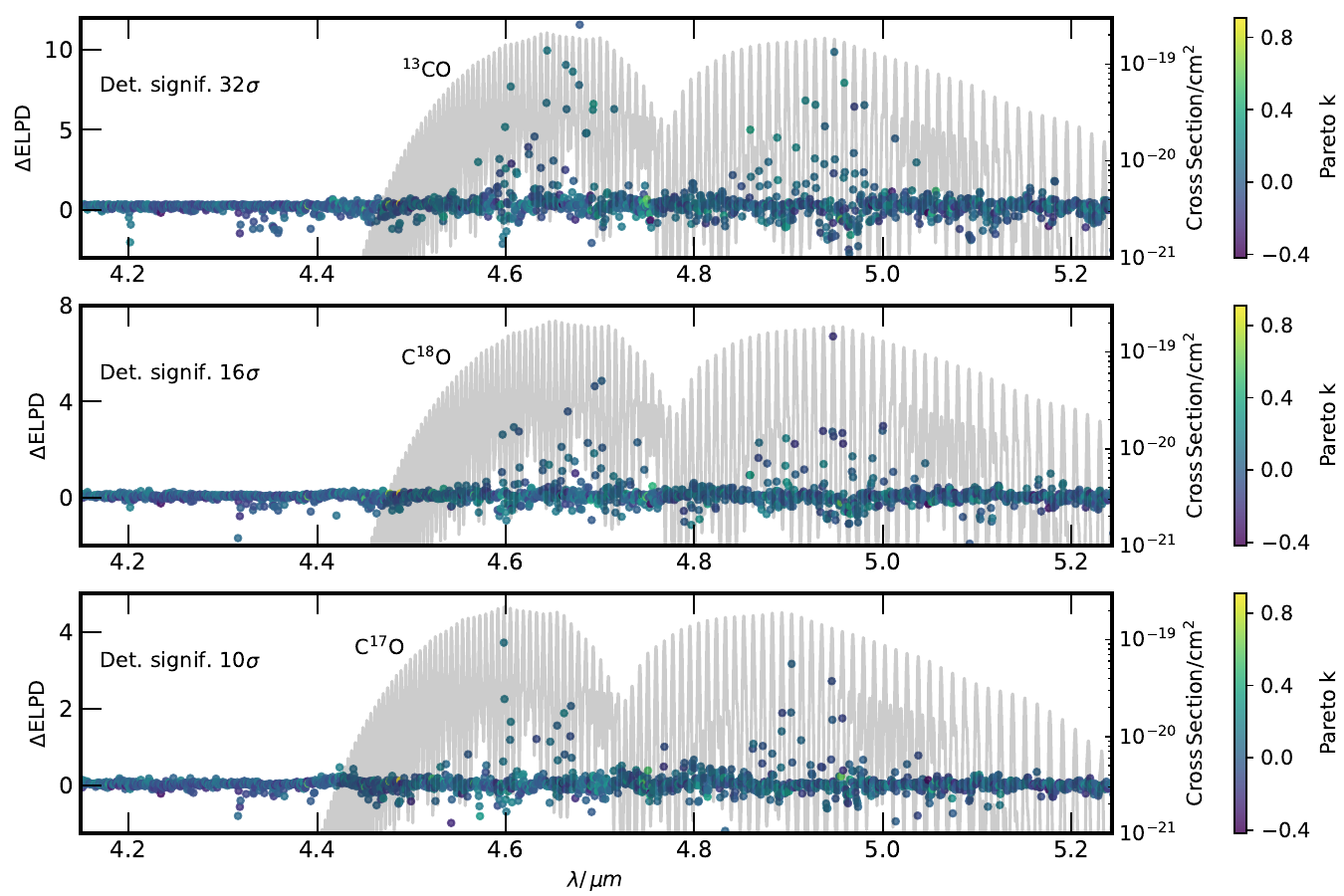}
    \caption{Difference in the expected log posterior predictive density (ELPD) for each isotopologue against our fiducial model. A positive $\Delta \mathrm{ELPD}$ score indicates that the model with the isotopologue is preferred by that data point. The colour bar indicates the Pareto k values for each of the data points, with points $\lesssim$ 0.7 indicating a good PSIS approximation (see section~\ref{sec:robustness}). The vast majority of the data points fall into this category and hence our use of the leave-one-out cross-validation is robust for all three isotopologues. In grey we also show the molecular cross section for each of the isotopologues at 1~bar pressure and 1200~K temperature at the spectral resolution of the observations (R$=$2,700).}     
\label{fig:loo}
\end{figure*}

\begin{figure}
\centering
	\includegraphics[width=\columnwidth,trim={0cm 0cm 0cm 0},clip]{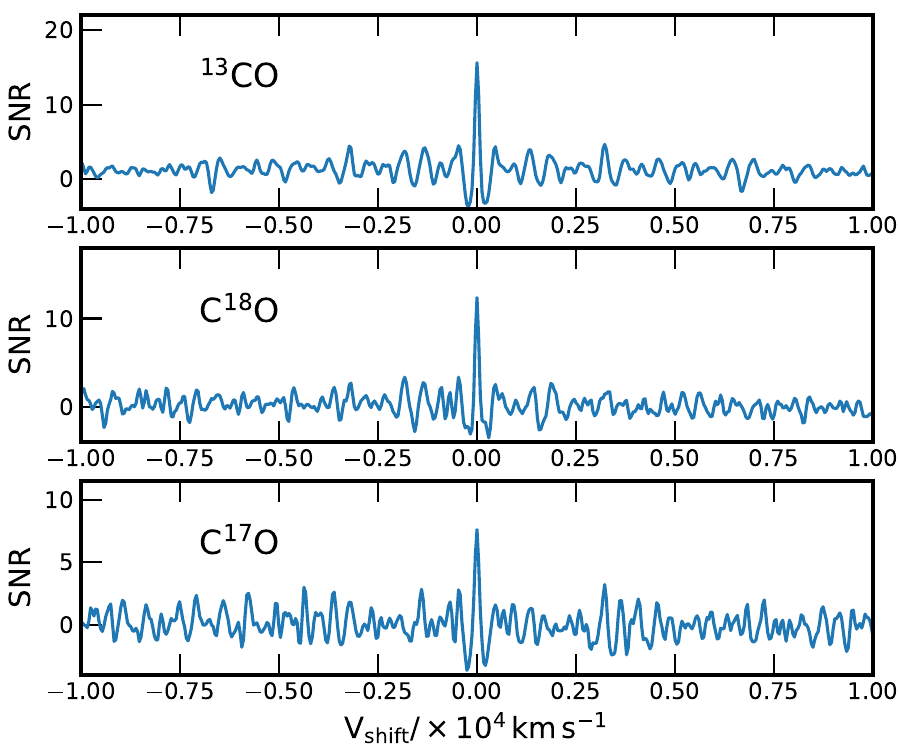}
    \caption{Signal-to-noise ratio (SNR) derived from the cross-correlation functions of the CO isotopologues.}     
\label{fig:CCF}
\end{figure}

Figure~\ref{fig:bestfit} shows the best fit retrieved spectrum to the observations and its residuals. Overall, this shows a good fit to the data, with the residuals only showing some low order deviations in the 4.8-5.1~$\mu$m range. The residuals may arise due to other species present in the atmosphere which are not accounted for in our retrieval, potentially even other isotopologues of molecules such as H$_2$O and CO$_2$. Generally however, the model shows a good fit, in particular for the clearly visible $^{12}$CO lines, as well as the H$_2$O and CO$_2$ features which are more prominent in the bluer regions of the spectrum, confirming the detections from \citet{miles2023}.

The retrieved P-T profile is shown in Figure~\ref{fig:PT}, and indicates a generally increasing temperature with pressure, as expected for non-irradiated objects. However, we do see that the temperature profile has a very slight inverted profile at $\sim$10~bar, which is near the edge of the photosphere where the continuum is set. This is likely because there may be additional species that we have not accounted for which is degenerate with the P-T profile of the atmosphere, for instance arising from other trace species or isotopes of CO$_2$ and H$_2$O. Overall though the temperature profile is consistent with expectations for such objects near the L/T transition and the photospheric temperature of $\sim$1100~K agrees well with previous works \citep{zhou2020, hoch2022, petrus2023, miles2023}.

\subsection{CO Isotopologues}\label{sec:co}

We constrain the chemical abundances of $^{12}$CO, the primary isotopologue of CO, as well as the minor isotopes $^{13}$CO, C$^{18}$O and C$^{17}$O, as shown in Figure~\ref{fig:posterior}. The isotopologues are distinguishable given that their opacities differ at the R$\sim$2,700 resolution of the observations. There is a strong interdependence between species, common from previous observations of such targets \citep[e.g.,][]{line2015, burningham2017, piette2020}. This highlights that the absolute abundances of each species may not be as well constrained, but we find that their ratios are significantly more precise and robust, as also seen with ground-based high-resolution spectroscopy \citep{gibson2022, gandhi2023, pelletier2023}. The less prominent isotopologues, C$^{18}$O and C$^{17}$O, show the weakest correlation given their lower abundances. This first detection of C$^{18}$O and C$^{17}$O in an exoplanet atmosphere demonstrates how the high quality JWST spectra can constrain minor isotopes.

\subsubsection{Robustness Tests}\label{sec:robustness}

We ensured that our constraints were robust through performing a Bayesian leave-one-out cross-validation \citep{welbanks2023}. This method fits a model to the full data with each data point removed by determining the expected log posterior predictive density (ELPD). Comparing with ELPD scores of a model with and without a given isotopologue therefore allows us to determine which data points most strongly prefer the isotopologue. Note that in principle a separate retrieval would need to be performed for every single data point, but we use the Pareto Smoothed Importance Sampling (PSIS) \citep{vehtari2017}. The Pareto k distribution fitted to the distribution of importance weights is a measure of how much the posterior changes with the removal of a single data point. Typically, Pareto k values $\lesssim$0.7 are a good approximation, but data points with values $>$0.7 will generally require a full retrieval without that point to fully assess the effect of removing it on the posterior. 

\begin{figure*}
\centering
	\includegraphics[width=\textwidth,trim={0cm 0cm 0cm 0},clip]{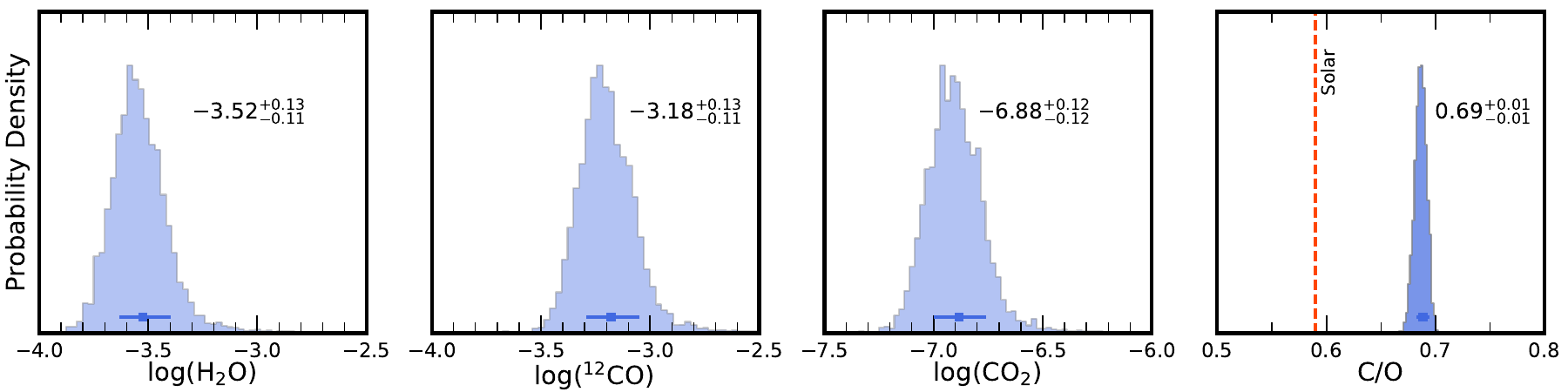}
    \caption{Marginalised posterior distributions for the volume mixing ratios of H$_2$O, $^{12}$CO, CO$_2$ and the overall atmospheric C/O ratio (right panel) from the retrieval of the $\sim$4.1-5.3~$\mu$m JWST observations of VHS~1256~b. The red dashed line shows the solar C/O ratio \citep{asplund2021}.}     
\label{fig:h2o_co_co2}
\end{figure*}

Figure~\ref{fig:loo} shows the $\Delta \mathrm{ELPD}$ and Pareto k values for each data point, with high $\Delta \mathrm{ELPD}$ values indicating that this data point prefers a model with the isotopologue. For all three of the isotopologues we find that the region of the spectrum where the detections are most influenced are the regions where each species has strong opacity. In addition, each isotopologue detection is driven by numerous data points, and the vast majority of the data points have Pareto k values $<$0.7, indicating a good approximation with PSIS. Therefore, each of our detections are robust. Note that not all of the wavelengths where the isotopologues have features have high $\Delta \mathrm{ELPD}$ values as other species such as H$_2$O, CO$_2$ and $^{12}$C$^{16}$O have stronger features and therefore interfere with the isotopologue signals at some wavelengths. We determine the detection significances of each isotopologue through comparison of the Bayesian evidence of retrievals performed without each. We detect $^{13}$CO, C$^{18}$O and C$^{17}$O very strongly, at significances of 32$\sigma$, 16$\sigma$ and 10$\sigma$ respectively. In addition, the error scale factor was increased when the isotopologues were not included in the retrieval, with the retrievals without $^{13}$CO, C$^{18}$O and C$^{17}$O showing increases of 37\%, 6\% and 4\% respectively. Hence, not including these species is a worse fit to the observations and thus requires a larger scaling of the error bars in order to match the model to the data.

We performed retrievals with a high pass filter applied to the data, which removes low order variability in the spectrum and ensures that the fit is performed on the spectral lines and their strengths only. This is particularly important to test as there is some low order deviations between the data and the best fit model in Figure~\ref{fig:bestfit}. To achieve this we used a running median of 20 data points on the observations and for every model in the retrieval. We found no significant change to any of the isotope ratios, and the detection significances remained similar, at 31$\sigma$, 15$\sigma$ and 10$\sigma$ for $^{13}$CO, C$^{18}$O and C$^{17}$O respectively.

In addition to these tests, we also detected each of the isotopologues through cross-correlation, as shown in Figure~\ref{fig:CCF}. \citet{esparza-borges2023} showed that this technique is applicable to JWST observations by detecting CO in the transiting exoplanet WASP-39~b with NIRSpec observations. We follow the procedure of \citet{zhang2021_bd}, using the data residuals (data subtracted by the best-fit model without the particular species) with a model spectral template, calculated by subtraction of the best-fit model without the species from the best-fit model with all species. Both the data residuals and the model template are then high-pass filtered and cross-correlated. We determine the signal-to-noise by dividing the cross-correlation by its standard deviation and subtracting off the mean. Our signal-to-noise ratios for all three isotopologues are lower than their detection significances, most likely due to aliasing with the periodic CO lines, but other factors such as the under-sampling of the peak signal due to the wavelength spacing of the data points may also reduce the overall signal. Nevertheless, the detections of $^{13}$CO and the oxygen isotopologues, C$^{18}$O and C$^{17}$O, remains clearly visible over the noise.

\subsection{Other molecular species}
In addition to CO and its various isotopologues, we also constrain H$_2$O and CO$_2$ (primary isotopologues only) given their strong opacity in this range, with their marginalised volume mixing ratios shown in Figure~\ref{fig:h2o_co_co2}. We detect H$_2$O with a lower abundance than the $^{12}$CO by $\sim0.34$~dex, and hence the atmospheric C/O ratio is $0.69^{+0.01}_{-0.01}$. This is consistent with previous works \citep{hoch2022, petrus2023}. Note that, as before, the ratio is better constrained than the absolute abundances of H$_2$O and CO. The C/O ratio is higher than the solar value of 0.59 \citep{asplund2021}, but some oxygen may have condensed out in silicates \citep{burrows1999}, which will raise the measured atmospheric C/O value.

\begin{figure*}
\centering
	\includegraphics[width=\textwidth,trim={0cm 0cm 0cm 0},clip]{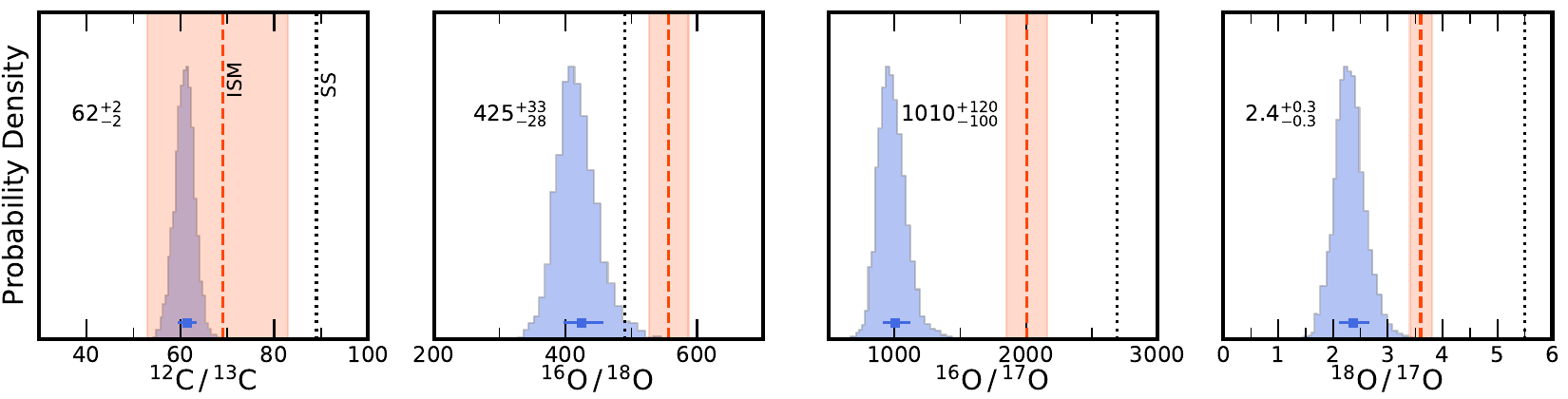}
    \caption{Isotope ratios $^{12}$C/$^{13}$C, $^{16}$O/$^{18}$O, $^{16}$O/$^{17}$O and $^{18}$O/$^{17}$O derived from the CO abundances of VHS~1256~b. The red dashed line and shaded area show the value measured for the local inter-stellar medium (ISM) and its 1$\sigma$ uncertainty \citep{wilson1999, milam2005}, and the dotted black line shows the value for the Solar System.}     
\label{fig:co}
\end{figure*}

The CO$_2$ abundance is much lower at $\log(\mathrm{CO_2}) = -6.88^{+0.12}_{-0.12}$, but its detection significance of 28$\sigma$ remains high given its very strong opacity at $\sim$4.2~$\mu$m. This confirms the results by \citet{miles2023}, who showed that the addition of CO$_2$ opacity improved the spectral fit to these JWST observations of VHS~1256~b. Carbon dioxide has been detected in a range of late-L to T-dwarfs \citep[e.g.,][]{sorahana2012} and is an excellent proxy for the metallicity, as it increases non-linearly with the overall metal content of the atmosphere in chemical equilibrium \citep[e.g.,][]{madhu2012, moses2013}. Our constrained abundance indicates an atmosphere that is at or near solar metallicity. We note however that the CO$_2$ abundance may be a strong function of the atmospheric pressure for such objects \citep{lodders2002}, and our assumption of a vertically independent volume mixing ratio may therefore break down if the vertical mixing is weak. 

We also retrieve the CH$_4$ and NH$_3$ abundances, but find only upper limits of $\log(\mathrm{CH_4}) < -6.2$ and $\log(\mathrm{NH_3}) < -6.8$ at 2$\sigma$ confidence as these species do not possess strong opacity a these wavelengths. Other wavelengths such as the $\sim$3-4~$\mu$m range have shown that CH$_4$ is present in the atmosphere, which has also indicated that the atmosphere is in chemical disequilibrium \citep{miles2018, miles2023}. However, the CH$_4$ is lower abundance than the CO and hence will not significantly alter C/O estimates.

\section{Discussion and Conclusions}

We have constrained the $^{13}$C, $^{18}$O and $^{17}$O isotopes from the CO abundances for the directly imaged companion VHS~1256~b through retrievals of the JWST NIRSpec G395H/F290LP observations in the 4.1-5.3~$\mu$m range \citep{miles2023}. Figure~\ref{fig:co} shows the isotope ratios for $^{12}$C/$^{13}$C, $^{16}$O/$^{18}$O, $^{16}$O/$^{17}$O and $^{18}$O/$^{17}$O. These ratios are more precise and robust than the absolute abundances, as discussed in section~\ref{sec:co}. All of the ratios have median values which are lower than for the ISM and Solar System, but the $^{12}$C/$^{13}$C ratio is still within the 1$\sigma$ uncertainty of the local ISM value \citep{milam2005}. The ratios are also lower than $^{12}$C/$^{13}$C and $^{16}$O/$^{18}$O values measured for an M-dwarf binary \citep{crossfield2019}. Our $^{12}$C/$^{13}$C value is in between the value of $\sim$30 obtained for planetary companions YSES-1~b and WASP-77~A~b \citep{zhang2021_nature, line2021} and the value of $\gtrsim$100 constrained for isolated brown dwarfs \citep{zhang2021_bd, deregt2023}.

Constraining the carbon and oxygen isotopes in an exoplanet atmosphere is key as isotope fractionation processes in protoplanetary disks can drive differences in their abundances. Several key processes can alter their ratios during planetary formation. Firstly, isotopologue selective photodissociation will prevent the more abundant isotopologues from dissociating because their higher column density will prevent stellar UV photons penetrating deep into a protoplanetary disk \citep[e.g.,][]{visser2009}. The less prominent isotopes will not be as effectively shielded and will therefore be more dissociated, resulting in the gas-phase ratios of more abundant isotopes to less abundant isotopes increasing. Secondly, the gas and ice phase interactions of CO can lead to isotopologue partitioning, as has been shown for $^{12}$C/$^{13}$C ratios \citep{smith2015}. Furthermore, isotope ion exchange reactions can increase the gas phase $^{13}$CO in protoplanetary disks, but generally does not have a strong impact on the oxygen isotopes \citep{langer1984}. A higher excitation temperature of $^{12}$CO than $^{13}$CO can also provide differences in the isotopologue ratio of these two species \citep{goto2003}.

Through this work we have been able to constrain the minor oxygen isotopes in a Super Jupiter for the first time. The $^{18}$O and $^{17}$O are lower abundance than $^{13}$C, but their ratios are still well constrained given the high precision and signal-to-noise observations. Both the $^{18}$O and $^{17}$O are more abundant than expected, with the $^{16}$O/$^{18}$O and $^{16}$O/$^{17}$O ratios lower than for the local ISM and Solar System values. This could be due to ices richer in the weaker isotopes being accreted during formation, potentially driven by isotope selective photodissociation. Any photodissociation of the gas phase would increase the ratio, which goes against our low retrieved ratios. A previous hint of C$^{18}$O in a brown dwarf indicated a $^{16}$O/$^{18}$O ratio nearer to $\sim$1500 \citep{zhang2022}, higher than our constrained value. This could be a potential indication of chemical inhomogeneity of molecular clouds between planetary systems. If protoplanetary disks form from evolved intermediate mass stars, the systems may be more enhanced in minor isotopes.

The ratio of the minor oxygen isotopes, $^{18}$O/$^{17}$O, is also lower, with our value more akin to that of the galactic centre. In addition to potential isotope fractionation processes, the enhanced $^{17}$O in VHS~1256~b may be an indication of accretion of material from lower mass stars, which are expected to be richer in this isotope \citep{wouterloot2008}. Note however that the $^{17}$O constraint from our work is the lowest abundance and the weakest detection significance, and therefore may not be as reliable as those for the other isotopes. In addition, VHS~1256~b shows significant time variability \citep{zhou2022}, which is most likely to affect our abundance constraints for the weaker species.

Our findings highlight the power of JWST to determine isotope ratios in exoplanetary atmospheres. The highly robust detections of both carbon and oxygen isotopologues of CO at very high significances demonstrate how we can expand atmospheric science into new avenues in planetary formation through isotopic chemistry. In the future, we may be able to extend this to determine D/H ratios for exoplanet atmospheres, which is an excellent tracer for planet formation \citep{morley2019} as well as the evolution of targets near the deuterium-burning mass boundary \citep{dupuy2023}. However, this is challenging as deuterium is typically $\sim10^4\times$ less abundant than $^1$H, and high accuracy and high temperature line lists for deuterium-rich species are required \citep[e.g.,][]{voronin2010}. We may additionally be able to constrain isotopic ratios for close-in transiting planets, thereby providing a contrast between hot Jupiters and the more widely separated planet populations. Another important insight will be comparison of the isotope ratios in host stars to determine whether differences in exoplanetary atmospheric ratios are caused by isotope fractionation or intrinsic differences in initial formation chemistry. This will also set the stage for the next generation of facilities such as the Extremely Large Telescope (ELT), with instruments such as METIS \citep{brandl2021} and ANDES \citep{maiolino2013} capable of very high precision and spectral resolution atmospheric observations in the coming decade.

\section*{Acknowledgements}
SG is grateful to Leiden Observatory at Leiden University for the award of the Oort Fellowship. This work was performed using the compute resources from the Academic Leiden Interdisciplinary Cluster Environment (ALICE) provided by Leiden University. We also utilise the Avon HPC cluster managed by the Scientific Computing Research Technology Platform (SCRTP) at the University of Warwick. SdR and IS acknowledge funding from the European Research Council (ERC) under the European Union’s Horizon 2020 research and innovation program under grant agreement No. 694513. We thank B. Miles and the JWST Early Release Science Programme for Direct Observations of Exoplanetary Systems for the NIRSpec data of VHS~1256~b. The JWST data presented in this paper were obtained from the Mikulski Archive for Space Telescopes (MAST) at the Space Telescope Science Institute. The specific observations analysed can be accessed via \dataset[10.17909/cv44-c816]{http://dx.doi.org/10.17909/cv44-c816}. For the purpose of open access, the author has applied a Creative Commons Attribution (CC-BY) licence to any Author Accepted Manuscript version arising from this submission. We thank the anonymous referee for a careful review of our manuscript.

%

\vspace{5mm}
\facilities{JWST (NIRSpec)}








\bibliography{refs}{}

\begin{thebibliography}{}
\expandafter\ifx\csname natexlab\endcsname\relax\def\natexlab#1{#1}\fi
\providecommand{\url}[1]{\href{#1}{#1}}
\providecommand{\dodoi}[1]{doi:~\href{http://doi.org/#1}{\nolinkurl{#1}}}
\providecommand{\doeprint}[1]{\href{http://ascl.net/#1}{\nolinkurl{http://ascl.net/#1}}}
\providecommand{\doarXiv}[1]{\href{https://arxiv.org/abs/#1}{\nolinkurl{https://arxiv.org/abs/#1}}}

\bibitem[{{Alderson} {et~al.}(2023){Alderson}, {Wakeford}, {Alam}, {Batalha},
  {Lothringer}, {Adams Redai}, {Barat}, {Brande}, {Damiano}, {Daylan},
  {Espinoza}, {Flagg}, {Goyal}, {Grant}, {Hu}, {Inglis}, {Lee}, {Mikal-Evans},
  {Ramos-Rosado}, {Roy}, {Wallack}, {Batalha}, {Bean}, {Benneke},
  {Berta-Thompson}, {Carter}, {Changeat}, {Col{\'o}n}, {Crossfield},
  {D{\'e}sert}, {Foreman-Mackey}, {Gibson}, {Kreidberg}, {Line},
  {L{\'o}pez-Morales}, {Molaverdikhani}, {Moran}, {Morello}, {Moses},
  {Mukherjee}, {Schlawin}, {Sing}, {Stevenson}, {Taylor}, {Aggarwal}, {Ahrer},
  {Allen}, {Barstow}, {Bell}, {Blecic}, {Casewell}, {Chubb}, {Crouzet},
  {Cubillos}, {Decin}, {Feinstein}, {Fortney}, {Harrington}, {Heng}, {Iro},
  {Kempton}, {Kirk}, {Knutson}, {Krick}, {Leconte}, {Lendl}, {MacDonald},
  {Mancini}, {Mansfield}, {May}, {Mayne}, {Miguel}, {Nikolov}, {Ohno}, {Palle},
  {Parmentier}, {Petit dit de la Roche}, {Piaulet}, {Powell}, {Rackham},
  {Redfield}, {Rogers}, {Rustamkulov}, {Tan}, {Tremblin}, {Tsai}, {Turner}, {de
  Val-Borro}, {Venot}, {Welbanks}, {Wheatley}, \& {Zhang}}]{anderson2023}
{Alderson}, L., {Wakeford}, H.~R., {Alam}, M.~K., {et~al.} 2023, \nat, 614,
  664, \dodoi{10.1038/s41586-022-05591-3}

\bibitem[{{Asplund} {et~al.}(2021){Asplund}, {Amarsi}, \&
  {Grevesse}}]{asplund2021}
{Asplund}, M., {Amarsi}, A.~M., \& {Grevesse}, N. 2021, \aap, 653, A141,
  \dodoi{10.1051/0004-6361/202140445}

\bibitem[{{Brandl} {et~al.}(2021){Brandl}, {Bettonvil}, {van Boekel},
  {Glauser}, {Quanz}, {Absil}, {Amorim}, {Feldt}, {Glasse}, {G{\"u}del}, {Ho},
  {Labadie}, {Meyer}, {Pantin}, {van Winckel}, \& {METIS
  Consortium}}]{brandl2021}
{Brandl}, B., {Bettonvil}, F., {van Boekel}, R., {et~al.} 2021, The Messenger,
  182, 22, \dodoi{10.18727/0722-6691/5218}

\bibitem[{{Buchner} {et~al.}(2014){Buchner}, {Georgakakis}, {Nandra}, {Hsu},
  {Rangel}, {Brightman}, {Merloni}, {Salvato}, {Donley}, \&
  {Kocevski}}]{buchner2014}
{Buchner}, J., {Georgakakis}, A., {Nandra}, K., {et~al.} 2014, \aap, 564, A125,
  \dodoi{10.1051/0004-6361/201322971}

\bibitem[{{Burningham} {et~al.}(2017){Burningham}, {Marley}, {Line}, {Lupu},
  {Visscher}, {Morley}, {Saumon}, \& {Freedman}}]{burningham2017}
{Burningham}, B., {Marley}, M.~S., {Line}, M.~R., {et~al.} 2017, \mnras, 470,
  1177, \dodoi{10.1093/mnras/stx1246}

\bibitem[{{Burrows} \& {Sharp}(1999)}]{burrows1999}
{Burrows}, A., \& {Sharp}, C.~M. 1999, \apj, 512, 843, \dodoi{10.1086/306811}

\bibitem[{{Chilcote} {et~al.}(2017){Chilcote}, {Pueyo}, {De Rosa}, {Vargas},
  {Macintosh}, {Bailey}, {Barman}, {Bauman}, {Bruzzone}, {Bulger}, {Burrows},
  {Cardwell}, {Chen}, {Cotten}, {Dillon}, {Doyon}, {Draper}, {Duch{\^e}ne},
  {Dunn}, {Erikson}, {Fitzgerald}, {Follette}, {Gavel}, {Goodsell}, {Graham},
  {Greenbaum}, {Hartung}, {Hibon}, {Hung}, {Ingraham}, {Kalas}, {Konopacky},
  {Larkin}, {Maire}, {Marchis}, {Marley}, {Marois}, {Metchev},
  {Millar-Blanchaer}, {Morzinski}, {Nielsen}, {Norton}, {Oppenheimer},
  {Palmer}, {Patience}, {Perrin}, {Poyneer}, {Rajan}, {Rameau},
  {Rantakyr{\"o}}, {Sadakuni}, {Saddlemyer}, {Savransky}, {Schneider}, {Serio},
  {Sivaramakrishnan}, {Song}, {Soummer}, {Thomas}, {Wallace}, {Wang},
  {Ward-Duong}, {Wiktorowicz}, \& {Wolff}}]{chilcote2017}
{Chilcote}, J., {Pueyo}, L., {De Rosa}, R.~J., {et~al.} 2017, \aj, 153, 182,
  \dodoi{10.3847/1538-3881/aa63e9}

\bibitem[{{Coles} {et~al.}(2019){Coles}, {Yurchenko}, \&
  {Tennyson}}]{coles2019}
{Coles}, P.~A., {Yurchenko}, S.~N., \& {Tennyson}, J. 2019, \mnras, 490, 4638,
  \dodoi{10.1093/mnras/stz2778}

\bibitem[{{Constantinou} {et~al.}(2023){Constantinou}, {Madhusudhan}, \&
  {Gandhi}}]{constantinou2023}
{Constantinou}, S., {Madhusudhan}, N., \& {Gandhi}, S. 2023, \apjl, 943, L10,
  \dodoi{10.3847/2041-8213/acaead}

\bibitem[{{Crossfield} {et~al.}(2019){Crossfield}, {Lothringer}, {Flores},
  {Mills}, {Freedman}, {Valverde}, {Miles}, {Guo}, \&
  {Skemer}}]{crossfield2019}
{Crossfield}, I.~J.~M., {Lothringer}, J.~D., {Flores}, B., {et~al.} 2019,
  \apjl, 871, L3, \dodoi{10.3847/2041-8213/aaf9b6}

\bibitem[{{de Regt} {et~al.}(2023){de Regt}, {Snellen}, \&
  {Gandhi}}]{deregt2023}
{de Regt}, S., {Snellen}, I., \& {Gandhi}, S. 2023, A\&A, \textit{in prep}

\bibitem[{{Dupuy} {et~al.}(2023){Dupuy}, {Liu}, {Evans}, {Best}, {Pearce},
  {Sanghi}, {Phillips}, \& {Bardalez Gagliuffi}}]{dupuy2023}
{Dupuy}, T.~J., {Liu}, M.~C., {Evans}, E.~L., {et~al.} 2023, \mnras, 519, 1688,
  \dodoi{10.1093/mnras/stac3557}

\bibitem[{{Esparza-Borges} {et~al.}(2023){Esparza-Borges}, {L{\'o}pez-Morales},
  {Adams Redai}, {Pall{\'e}}, {Kirk}, {Casasayas-Barris}, {Batalha}, {Rackham},
  {Bean}, {Casewell}, {Decin}, {Dos Santos}, {Garc{\'\i}a Mu{\~n}oz},
  {Harrington}, {Heng}, {Hu}, {Mancini}, {Molaverdikhani}, {Morello},
  {Nikolov}, {Nixon}, {Redfield}, {Stevenson}, {Wakeford}, {Alam}, {Benneke},
  {Blecic}, {Crouzet}, {Daylan}, {Inglis}, {Kreidberg}, {Petit dit de la
  Roche}, \& {Turner}}]{esparza-borges2023}
{Esparza-Borges}, E., {L{\'o}pez-Morales}, M., {Adams Redai}, J.~I., {et~al.}
  2023, \apjl, 955, L19, \dodoi{10.3847/2041-8213/acf27b}

\bibitem[{{Feroz} \& {Hobson}(2008)}]{feroz2008}
{Feroz}, F., \& {Hobson}, M.~P. 2008, \mnras, 384, 449,
  \dodoi{10.1111/j.1365-2966.2007.12353.x}

\bibitem[{{Feroz} {et~al.}(2009){Feroz}, {Hobson}, \& {Bridges}}]{feroz2009}
{Feroz}, F., {Hobson}, M.~P., \& {Bridges}, M. 2009, \mnras, 398, 1601,
  \dodoi{10.1111/j.1365-2966.2009.14548.x}

\bibitem[{{Gandhi} \& {Madhusudhan}(2018)}]{gandhi2018}
{Gandhi}, S., \& {Madhusudhan}, N. 2018, \mnras, 474, 271,
  \dodoi{10.1093/mnras/stx2748}

\bibitem[{{Gandhi} {et~al.}(2020){Gandhi}, {Brogi}, {Yurchenko}, {Tennyson},
  {Coles}, {Webb}, {Birkby}, {Guilluy}, {Hawker}, {Madhusudhan}, {Bonomo}, \&
  {Sozzetti}}]{gandhi2020_cs}
{Gandhi}, S., {Brogi}, M., {Yurchenko}, S.~N., {et~al.} 2020, \mnras, 495, 224,
  \dodoi{10.1093/mnras/staa981}

\bibitem[{{Gandhi} {et~al.}(2023){Gandhi}, {Kesseli}, {Zhang}, {Louca},
  {Snellen}, {Brogi}, {Miguel}, {Casasayas-Barris}, {Pelletier}, {Landman},
  {Maguire}, \& {Gibson}}]{gandhi2023}
{Gandhi}, S., {Kesseli}, A., {Zhang}, Y., {et~al.} 2023, \aj, 165, 242,
  \dodoi{10.3847/1538-3881/accd65}

\bibitem[{{Gaudi} {et~al.}(2020){Gaudi}, {Seager}, {Mennesson}, {Kiessling},
  {Warfield}, {Cahoy}, {Clarke}, {Domagal-Goldman}, {Feinberg}, {Guyon},
  {Kasdin}, {Mawet}, {Plavchan}, {Robinson}, {Rogers}, {Scowen}, {Somerville},
  {Stapelfeldt}, {Stark}, {Stern}, {Turnbull}, {Amini}, {Kuan}, {Martin},
  {Morgan}, {Redding}, {Stahl}, {Webb}, {Alvarez-Salazar}, {Arnold}, {Arya},
  {Balasubramanian}, {Baysinger}, {Bell}, {Below}, {Benson}, {Blais}, {Booth},
  {Bourgeois}, {Bradford}, {Brewer}, {Brooks}, {Cady}, {Caldwell}, {Calvet},
  {Carr}, {Chan}, {Cormarkovic}, {Coste}, {Cox}, {Danner}, {Davis}, {Dewell},
  {Dorsett}, {Dunn}, {East}, {Effinger}, {Eng}, {Freebury}, {Garcia}, {Gaskin},
  {Greene}, {Hennessy}, {Hilgemann}, {Hood}, {Holota}, {Howe}, {Huang}, {Hull},
  {Hunt}, {Hurd}, {Johnson}, {Kissil}, {Knight}, {Kolenz}, {Kraus}, {Krist},
  {Li}, {Lisman}, {Mandic}, {Mann}, {Marchen}, {Marrese-Reading}, {McCready},
  {McGown}, {Missun}, {Miyaguchi}, {Moore}, {Nemati}, {Nikzad}, {Nissen},
  {Novicki}, {Perrine}, {Pineda}, {Polanco}, {Putnam}, {Qureshi}, {Richards},
  {Eldorado Riggs}, {Rodgers}, {Rud}, {Saini}, {Scalisi}, {Scharf}, {Schulz},
  {Serabyn}, {Sigrist}, {Sikkia}, {Singleton}, {Shaklan}, {Smith}, {Southerd},
  {Stahl}, {Steeves}, {Sturges}, {Sullivan}, {Tang}, {Taras}, {Tesch},
  {Therrell}, {Tseng}, {Valente}, {Van Buren}, {Villalvazo}, {Warwick}, {Webb},
  {Westerhoff}, {Wofford}, {Wu}, {Woo}, {Wood}, {Ziemer}, {Arney}, {Anderson},
  {Ma{\'\i}z-Apell{\'a}niz}, {Bartlett}, {Belikov}, {Bendek}, {Cenko},
  {Douglas}, {Dulz}, {Evans}, {Faramaz}, {Feng}, {Ferguson}, {Follette},
  {Ford}, {Garc{\'\i}a}, {Geha}, {Gelino}, {G{\"o}tberg}, {Hildebrandt}, {Hu},
  {Jahnke}, {Kennedy}, {Kreidberg}, {Isella}, {Lopez}, {Marchis}, {Macri},
  {Marley}, {Matzko}, {Mazoyer}, {McCandliss}, {Meshkat}, {Mordasini},
  {Morris}, {Nielsen}, {Newman}, {Petigura}, {Postman}, {Reines}, {Roberge},
  {Roederer}, {Ruane}, {Schwieterman}, {Sirbu}, {Spalding}, {Teplitz},
  {Tumlinson}, {Turner}, {Werk}, {Wofford}, {Wyatt}, {Young}, \&
  {Zellem}}]{gaudi2020}
{Gaudi}, B.~S., {Seager}, S., {Mennesson}, B., {et~al.} 2020, arXiv e-prints,
  arXiv:2001.06683, \dodoi{10.48550/arXiv.2001.06683}

\bibitem[{{Gauza} {et~al.}(2015){Gauza}, {B{\'e}jar}, {P{\'e}rez-Garrido},
  {Zapatero Osorio}, {Lodieu}, {Rebolo}, {Pall{\'e}}, \& {Nowak}}]{gauza2015}
{Gauza}, B., {B{\'e}jar}, V. J.~S., {P{\'e}rez-Garrido}, A., {et~al.} 2015,
  \apj, 804, 96, \dodoi{10.1088/0004-637X/804/2/96}

\bibitem[{{Gibson} {et~al.}(2022){Gibson}, {Nugroho}, {Lothringer}, {Maguire},
  \& {Sing}}]{gibson2022}
{Gibson}, N.~P., {Nugroho}, S.~K., {Lothringer}, J., {Maguire}, C., \& {Sing},
  D.~K. 2022, \mnras, 512, 4618, \dodoi{10.1093/mnras/stac091}

\bibitem[{{Goto} {et~al.}(2003){Goto}, {Usuda}, {Takato}, {Hayashi},
  {Sakamoto}, {Gaessler}, {Hayano}, {Iye}, {Kamata}, {Kanzawa}, {Kobayashi},
  {Minowa}, {Nedachi}, {Oya}, {Pyo}, {Saint-Jacques}, {Suto}, {Takami},
  {Terada}, \& {Mitchell}}]{goto2003}
{Goto}, M., {Usuda}, T., {Takato}, N., {et~al.} 2003, \apj, 598, 1038,
  \dodoi{10.1086/378978}

\bibitem[{{Grant} {et~al.}(2023){Grant}, {Lothringer}, {Wakeford}, {Alam},
  {Alderson}, {Bean}, {Benneke}, {D{\'e}sert}, {Daylan}, {Flagg}, {Hu},
  {Inglis}, {Kirk}, {Kreidberg}, {L{\'o}pez-Morales}, {Mancini}, {Mikal-Evans},
  {Molaverdikhani}, {Palle}, {Rackham}, {Redfield}, {Stevenson}, {Valenti},
  {Wallack}, {Aggarwal}, {Ahrer}, {Crossfield}, {Crouzet}, {Iro}, {Nikolov},
  {Wheatley}, \& {JWST Transiting Exoplanet Community ERS Team}}]{grant2023}
{Grant}, D., {Lothringer}, J.~D., {Wakeford}, H.~R., {et~al.} 2023, \apjl, 949,
  L15, \dodoi{10.3847/2041-8213/acd544}

\bibitem[{{Hargreaves} {et~al.}(2020){Hargreaves}, {Gordon}, {Rey}, {Nikitin},
  {Tyuterev}, {Kochanov}, \& {Rothman}}]{hargreaves2020}
{Hargreaves}, R.~J., {Gordon}, I.~E., {Rey}, M., {et~al.} 2020, \apjs, 247, 55,
  \dodoi{10.3847/1538-4365/ab7a1a}

\bibitem[{{Hoch} {et~al.}(2022){Hoch}, {Konopacky}, {Barman}, {Theissen},
  {Brock}, {Perrin}, {Ruffio}, {Macintosh}, \& {Marois}}]{hoch2022}
{Hoch}, K. K.~W., {Konopacky}, Q.~M., {Barman}, T.~S., {et~al.} 2022, \aj, 164,
  155, \dodoi{10.3847/1538-3881/ac84d4}

\bibitem[{{Huang} {et~al.}(2013){Huang}, {Freedman}, {Tashkun}, {Schwenke}, \&
  {Lee}}]{huang2013}
{Huang}, X., {Freedman}, R.~S., {Tashkun}, S.~A., {Schwenke}, D.~W., \& {Lee},
  T.~J. 2013, \jqsrt, 130, 134, \dodoi{10.1016/j.jqsrt.2013.05.018}

\bibitem[{Huang {et~al.}(2017)Huang, Schwenke, Freedman, \& Lee}]{huang2017}
Huang, X., Schwenke, D.~W., Freedman, R.~S., \& Lee, T.~J. 2017, Journal of
  Quantitative Spectroscopy and Radiative Transfer, 203, 224 ,
  \dodoi{https://doi.org/10.1016/j.jqsrt.2017.04.026}

\bibitem[{{Janson} {et~al.}(2013){Janson}, {Brandt}, {Kuzuhara}, {Spiegel},
  {Thalmann}, {Currie}, {Bonnefoy}, {Zimmerman}, {Sorahana}, {Kotani},
  {Schlieder}, {Hashimoto}, {Kudo}, {Kusakabe}, {Abe}, {Brandner}, {Carson},
  {Egner}, {Feldt}, {Goto}, {Grady}, {Guyon}, {Hayano}, {Hayashi}, {Hayashi},
  {Henning}, {Hodapp}, {Ishii}, {Iye}, {Kandori}, {Knapp}, {Kwon}, {Matsuo},
  {McElwain}, {Mede}, {Miyama}, {Morino}, {Moro-Mart{\'\i}n}, {Nakagawa},
  {Nishimura}, {Pyo}, {Serabyn}, {Suenaga}, {Suto}, {Suzuki}, {Takahashi},
  {Takami}, {Takato}, {Terada}, {Tomono}, {Turner}, {Watanabe}, {Wisniewski},
  {Yamada}, {Takami}, {Usuda}, \& {Tamura}}]{janson2013}
{Janson}, M., {Brandt}, T.~D., {Kuzuhara}, M., {et~al.} 2013, \apjl, 778, L4,
  \dodoi{10.1088/2041-8205/778/1/L4}

\bibitem[{{JWST Transiting Exoplanet Community Early Release Science Team}
  {et~al.}(2023){JWST Transiting Exoplanet Community Early Release Science
  Team}, {Ahrer}, {Alderson}, {Batalha}, {Batalha}, {Bean}, {Beatty}, {Bell},
  {Benneke}, {Berta-Thompson}, {Carter}, {Crossfield}, {Espinoza}, {Feinstein},
  {Fortney}, {Gibson}, {Goyal}, {Kempton}, {Kirk}, {Kreidberg},
  {L{\'o}pez-Morales}, {Line}, {Lothringer}, {Moran}, {Mukherjee}, {Ohno},
  {Parmentier}, {Piaulet}, {Rustamkulov}, {Schlawin}, {Sing}, {Stevenson},
  {Wakeford}, {Allen}, {Birkmann}, {Brande}, {Crouzet}, {Cubillos}, {Damiano},
  {D{\'e}sert}, {Gao}, {Harrington}, {Hu}, {Kendrew}, {Knutson}, {Lagage},
  {Leconte}, {Lendl}, {MacDonald}, {May}, {Miguel}, {Molaverdikhani}, {Moses},
  {Murray}, {Nehring}, {Nikolov}, {Petit dit de la Roche}, {Radica}, {Roy},
  {Stassun}, {Taylor}, {Waalkes}, {Wachiraphan}, {Welbanks}, {Wheatley},
  {Aggarwal}, {Alam}, {Banerjee}, {Barstow}, {Blecic}, {Casewell}, {Changeat},
  {Chubb}, {Col{\'o}n}, {Coulombe}, {Daylan}, {de Val-Borro}, {Decin}, {Dos
  Santos}, {Flagg}, {France}, {Fu}, {Garc{\'\i}a Mu{\~n}oz}, {Gizis},
  {Glidden}, {Grant}, {Heng}, {Henning}, {Hong}, {Inglis}, {Iro}, {Kataria},
  {Komacek}, {Krick}, {Lee}, {Lewis}, {Lillo-Box}, {Lustig-Yaeger}, {Mancini},
  {Mandell}, {Mansfield}, {Marley}, {Mikal-Evans}, {Morello}, {Nixon}, {Ortiz
  Ceballos}, {Piette}, {Powell}, {Rackham}, {Ramos-Rosado}, {Rauscher},
  {Redfield}, {Rogers}, {Roman}, {Roudier}, {Scarsdale}, {Shkolnik},
  {Southworth}, {Spake}, {Steinrueck}, {Tan}, {Teske}, {Tremblin}, {Tsai},
  {Tucker}, {Turner}, {Valenti}, {Venot}, {Waldmann}, {Wallack}, {Zhang}, \&
  {Zieba}}]{jwst2023}
{JWST Transiting Exoplanet Community Early Release Science Team}, {Ahrer},
  E.-M., {Alderson}, L., {et~al.} 2023, \nat, 614, 649,
  \dodoi{10.1038/s41586-022-05269-w}

\bibitem[{{Langer} {et~al.}(1984){Langer}, {Graedel}, {Frerking}, \&
  {Armentrout}}]{langer1984}
{Langer}, W.~D., {Graedel}, T.~E., {Frerking}, M.~A., \& {Armentrout}, P.~B.
  1984, \apj, 277, 581, \dodoi{10.1086/161730}

\bibitem[{{Lavie} {et~al.}(2017){Lavie}, {Mendon{\c{c}}a}, {Mordasini},
  {Malik}, {Bonnefoy}, {Demory}, {Oreshenko}, {Grimm}, {Ehrenreich}, \&
  {Heng}}]{lavie2017}
{Lavie}, B., {Mendon{\c{c}}a}, J.~M., {Mordasini}, C., {et~al.} 2017, \aj, 154,
  91, \dodoi{10.3847/1538-3881/aa7ed8}

\bibitem[{{Lee} {et~al.}(2013){Lee}, {Heng}, \& {Irwin}}]{lee2013}
{Lee}, J.-M., {Heng}, K., \& {Irwin}, P. G.~J. 2013, \apj, 778, 97,
  \dodoi{10.1088/0004-637X/778/2/97}

\bibitem[{{Li} {et~al.}(2015){Li}, {Gordon}, {Rothman}, {Tan}, {Hu}, {Kassi},
  {Campargue}, \& {Medvedev}}]{li2015}
{Li}, G., {Gordon}, I.~E., {Rothman}, L.~S., {et~al.} 2015, \apjs, 216, 15,
  \dodoi{10.1088/0067-0049/216/1/15}

\bibitem[{{Li} \& {Cao}(2022)}]{li2022}
{Li}, Z., \& {Cao}, J. 2022, arXiv e-prints, arXiv:2201.06808,
  \dodoi{10.48550/arXiv.2201.06808}

\bibitem[{{Line} {et~al.}(2015){Line}, {Teske}, {Burningham}, {Fortney}, \&
  {Marley}}]{line2015}
{Line}, M.~R., {Teske}, J., {Burningham}, B., {Fortney}, J.~J., \& {Marley},
  M.~S. 2015, \apj, 807, 183, \dodoi{10.1088/0004-637X/807/2/183}

\bibitem[{{Line} {et~al.}(2021){Line}, {Brogi}, {Bean}, {Gandhi}, {Zalesky},
  {Parmentier}, {Smith}, {Mace}, {Mansfield}, {Kempton}, {Fortney}, {Shkolnik},
  {Patience}, {Rauscher}, {D{\'e}sert}, \& {Wardenier}}]{line2021}
{Line}, M.~R., {Brogi}, M., {Bean}, J.~L., {et~al.} 2021, \nat, 598, 580,
  \dodoi{10.1038/s41586-021-03912-6}

\bibitem[{{Lodders} \& {Fegley}(2002)}]{lodders2002}
{Lodders}, K., \& {Fegley}, B. 2002, \icarus, 155, 393,
  \dodoi{10.1006/icar.2001.6740}

\bibitem[{{Madhusudhan}(2012)}]{madhu2012}
{Madhusudhan}, N. 2012, \apj, 758, 36, \dodoi{10.1088/0004-637X/758/1/36}

\bibitem[{{Maiolino} {et~al.}(2013){Maiolino}, {Haehnelt}, {Murphy}, {Queloz},
  {Origlia}, {Alcala}, {Alibert}, {Amado}, {Allende Prieto}, {Ammler-von Eiff},
  {Asplund}, {Barstow}, {Becker}, {Bonfils}, {Bouchy}, {Bragaglia}, {Burleigh},
  {Chiavassa}, {Cimatti}, {Cirasuolo}, {Cristiani}, {D'Odorico}, {Dravins},
  {Emsellem}, {Farihi}, {Figueira}, {Fynbo}, {Gansicke}, {Gillon},
  {Gustafsson}, {Hill}, {Israelyan}, {Korn}, {Larsen}, {De Laverny}, {Liske},
  {Lovis}, {Marconi}, {Martins}, {Molaro}, {Nisini}, {Oliva}, {Petitjean},
  {Pettini}, {Recio Blanco}, {Rebolo}, {Reiners}, {Rodriguez-Lopez}, {Ryde},
  {Santos}, {Savaglio}, {Snellen}, {Strassmeier}, {Tanvir}, {Testi}, {Tolstoy},
  {Triaud}, {Vanzi}, {Viel}, \& {Volonteri}}]{maiolino2013}
{Maiolino}, R., {Haehnelt}, M., {Murphy}, M.~T., {et~al.} 2013, arXiv e-prints,
  arXiv:1310.3163, \dodoi{10.48550/arXiv.1310.3163}

\bibitem[{{Milam} {et~al.}(2005){Milam}, {Savage}, {Brewster}, {Ziurys}, \&
  {Wyckoff}}]{milam2005}
{Milam}, S.~N., {Savage}, C., {Brewster}, M.~A., {Ziurys}, L.~M., \& {Wyckoff},
  S. 2005, \apj, 634, 1126, \dodoi{10.1086/497123}

\bibitem[{{Miles} {et~al.}(2018){Miles}, {Skemer}, {Barman}, {Allers}, \&
  {Stone}}]{miles2018}
{Miles}, B.~E., {Skemer}, A.~J., {Barman}, T.~S., {Allers}, K.~N., \& {Stone},
  J.~M. 2018, \apj, 869, 18, \dodoi{10.3847/1538-4357/aae6cd}

\bibitem[{{Miles} {et~al.}(2023){Miles}, {Biller}, {Patapis}, {Worthen},
  {Rickman}, {Hoch}, {Skemer}, {Perrin}, {Whiteford}, {Chen}, {Sargent},
  {Mukherjee}, {Morley}, {Moran}, {Bonnefoy}, {Petrus}, {Carter}, {Choquet},
  {Hinkley}, {Ward-Duong}, {Leisenring}, {Millar-Blanchaer}, {Pueyo}, {Ray},
  {Sallum}, {Stapelfeldt}, {Stone}, {Wang}, {Absil}, {Balmer}, {Boccaletti},
  {Bonavita}, {Booth}, {Bowler}, {Chauvin}, {Christiaens}, {Currie},
  {Danielski}, {Fortney}, {Girard}, {Grady}, {Greenbaum}, {Henning}, {Hines},
  {Janson}, {Kalas}, {Kammerer}, {Kennedy}, {Kenworthy}, {Kervella}, {Lagage},
  {Lew}, {Liu}, {Macintosh}, {Marino}, {Marley}, {Marois}, {Matthews},
  {Matthews}, {Mawet}, {McElwain}, {Metchev}, {Meyer}, {Molliere}, {Pantin},
  {Quirrenbach}, {Rebollido}, {Ren}, {Schneider}, {Vasist}, {Wyatt}, {Zhou},
  {Briesemeister}, {Bryan}, {Calissendorff}, {Cantalloube}, {Cugno}, {De
  Furio}, {Dupuy}, {Factor}, {Faherty}, {Fitzgerald}, {Franson}, {Gonzales},
  {Hood}, {Howe}, {Kraus}, {Kuzuhara}, {Lagrange}, {Lawson}, {Lazzoni}, {Liu},
  {Llop-Sayson}, {Lloyd}, {Martinez}, {Mazoyer}, {Quanz}, {Redai}, {Samland},
  {Schlieder}, {Tamura}, {Tan}, {Uyama}, {Vigan}, {Vos}, {Wagner}, {Wolff},
  {Ygouf}, {Zhang}, {Zhang}, \& {Zhang}}]{miles2023}
{Miles}, B.~E., {Biller}, B.~A., {Patapis}, P., {et~al.} 2023, \apjl, 946, L6,
  \dodoi{10.3847/2041-8213/acb04a}

\bibitem[{{Molli{\`e}re} \& {Snellen}(2019)}]{molliere2019}
{Molli{\`e}re}, P., \& {Snellen}, I.~A.~G. 2019, \aap, 622, A139,
  \dodoi{10.1051/0004-6361/201834169}

\bibitem[{{Molli{\`e}re} {et~al.}(2020){Molli{\`e}re}, {Stolker}, {Lacour},
  {Otten}, {Shangguan}, {Charnay}, {Molyarova}, {Nowak}, {Henning}, {Marleau},
  {Semenov}, {van Dishoeck}, {Eisenhauer}, {Garcia}, {Garcia Lopez}, {Girard},
  {Greenbaum}, {Hinkley}, {Kervella}, {Kreidberg}, {Maire}, {Nasedkin},
  {Pueyo}, {Snellen}, {Vigan}, {Wang}, {de Zeeuw}, \& {Zurlo}}]{molliere2020}
{Molli{\`e}re}, P., {Stolker}, T., {Lacour}, S., {et~al.} 2020, \aap, 640,
  A131, \dodoi{10.1051/0004-6361/202038325}

\bibitem[{{Morley} {et~al.}(2019){Morley}, {Skemer}, {Miles}, {Line}, {Lopez},
  {Brogi}, {Freedman}, \& {Marley}}]{morley2019}
{Morley}, C.~V., {Skemer}, A.~J., {Miles}, B.~E., {et~al.} 2019, \apjl, 882,
  L29, \dodoi{10.3847/2041-8213/ab3c65}

\bibitem[{{Moses} {et~al.}(2013){Moses}, {Madhusudhan}, {Visscher}, \&
  {Freedman}}]{moses2013}
{Moses}, J.~I., {Madhusudhan}, N., {Visscher}, C., \& {Freedman}, R.~S. 2013,
  \apj, 763, 25, \dodoi{10.1088/0004-637X/763/1/25}

\bibitem[{{National Academies of Sciences, Engineering, and
  Medicine}(2021)}]{decadalsurvey2021}
{National Academies of Sciences, Engineering, and Medicine}. 2021, Pathways to
  Discovery in Astronomy and Astrophysics for the 2020s (Washington, DC: The
  National Academies Press), \dodoi{10.17226/26141}

\bibitem[{{Pelletier} {et~al.}(2023){Pelletier}, {Benneke}, {Ali-Dib},
  {Prinoth}, {Kasper}, {Seifahrt}, {Bean}, {Debras}, {Klein}, {Bazinet},
  {Hoeijmakers}, {Kesseli}, {Lim}, {Carmona}, {Pino}, {Casasayas-Barris},
  {Hood}, \& {St{\"u}rmer}}]{pelletier2023}
{Pelletier}, S., {Benneke}, B., {Ali-Dib}, M., {et~al.} 2023, \nat, 619, 491,
  \dodoi{10.1038/s41586-023-06134-0}

\bibitem[{{Petrus} {et~al.}(2023){Petrus}, {Chauvin}, {Bonnefoy}, {Tremblin},
  {Charnay}, {Delorme}, {Marleau}, {Bayo}, {Manjavacas}, {Lagrange},
  {Molli{\`e}re}, {Palma-Bifani}, {Biller}, {Jenkins}, {Goyal}, \&
  {Hoch}}]{petrus2023}
{Petrus}, S., {Chauvin}, G., {Bonnefoy}, M., {et~al.} 2023, \aap, 670, L9,
  \dodoi{10.1051/0004-6361/202244494}

\bibitem[{{Piette} \& {Madhusudhan}(2020)}]{piette2020}
{Piette}, A. A.~A., \& {Madhusudhan}, N. 2020, \mnras, 497, 5136,
  \dodoi{10.1093/mnras/staa2289}

\bibitem[{{Polyansky} {et~al.}(2018){Polyansky}, {Kyuberis}, {Zobov},
  {Tennyson}, {Yurchenko}, \& {Lodi}}]{polyansky2018}
{Polyansky}, O.~L., {Kyuberis}, A.~A., {Zobov}, N.~F., {et~al.} 2018, \mnras,
  480, 2597, \dodoi{10.1093/mnras/sty1877}

\bibitem[{{Quanz} {et~al.}(2022){Quanz}, {Ottiger}, {Fontanet}, {Kammerer},
  {Menti}, {Dannert}, {Gheorghe}, {Absil}, {Airapetian}, {Alei}, {Allart},
  {Angerhausen}, {Blumenthal}, {Buchhave}, {Cabrera},
  {Carri{\'o}n-Gonz{\'a}lez}, {Chauvin}, {Danchi}, {Dandumont}, {Defr{\'e}re},
  {Dorn}, {Ehrenreich}, {Ertel}, {Fridlund}, {Garc{\'\i}a Mu{\~n}oz},
  {Gasc{\'o}n}, {Girard}, {Glauser}, {Grenfell}, {Guidi}, {Hagelberg},
  {Helled}, {Ireland}, {Janson}, {Kopparapu}, {Korth}, {Kozakis}, {Kraus},
  {L{\'e}ger}, {Leedj{\"a}rv}, {Lichtenberg}, {Lillo-Box}, {Linz}, {Liseau},
  {Loicq}, {Mahendra}, {Malbet}, {Mathew}, {Mennesson}, {Meyer}, {Mishra},
  {Molaverdikhani}, {Noack}, {Oza}, {Pall{\'e}}, {Parviainen}, {Quirrenbach},
  {Rauer}, {Ribas}, {Rice}, {Romagnolo}, {Rugheimer}, {Schwieterman},
  {Serabyn}, {Sharma}, {Stassun}, {Szul{\'a}gyi}, {Wang}, {Wunderlich},
  {Wyatt}, \& {LIFE Collaboration}}]{quanz2022}
{Quanz}, S.~P., {Ottiger}, M., {Fontanet}, E., {et~al.} 2022, \aap, 664, A21,
  \dodoi{10.1051/0004-6361/202140366}

\bibitem[{{Richard} {et~al.}(2012){Richard}, {Gordon}, {Rothman}, {Abel},
  {Frommhold}, {Gustafsson}, {Hartmann}, {Hermans}, {Lafferty}, {Orton},
  {Smith}, \& {Tran}}]{richard2012}
{Richard}, C., {Gordon}, I.~E., {Rothman}, L.~S., {et~al.} 2012, \jqsrt, 113,
  1276, \dodoi{10.1016/j.jqsrt.2011.11.004}

\bibitem[{{Rothman} {et~al.}(2010){Rothman}, {Gordon}, {Barber}, {Dothe},
  {Gamache}, {Goldman}, {Perevalov}, {Tashkun}, \& {Tennyson}}]{rothman2010}
{Rothman}, L.~S., {Gordon}, I.~E., {Barber}, R.~J., {et~al.} 2010, \jqsrt, 111,
  2139, \dodoi{10.1016/j.jqsrt.2010.05.001}

\bibitem[{{Samland} {et~al.}(2017){Samland}, {Molli{\`e}re}, {Bonnefoy},
  {Maire}, {Cantalloube}, {Cheetham}, {Mesa}, {Gratton}, {Biller}, {Wahhaj},
  {Bouwman}, {Brandner}, {Melnick}, {Carson}, {Janson}, {Henning}, {Homeier},
  {Mordasini}, {Langlois}, {Quanz}, {van Boekel}, {Zurlo}, {Schlieder},
  {Avenhaus}, {Beuzit}, {Boccaletti}, {Bonavita}, {Chauvin}, {Claudi}, {Cudel},
  {Desidera}, {Feldt}, {Fusco}, {Galicher}, {Kopytova}, {Lagrange}, {Le
  Coroller}, {Martinez}, {Moeller-Nilsson}, {Mouillet}, {Mugnier}, {Perrot},
  {Sevin}, {Sissa}, {Vigan}, \& {Weber}}]{samland2017}
{Samland}, M., {Molli{\`e}re}, P., {Bonnefoy}, M., {et~al.} 2017, \aap, 603,
  A57, \dodoi{10.1051/0004-6361/201629767}

\bibitem[{{Smith} {et~al.}(2015){Smith}, {Pontoppidan}, {Young}, \&
  {Morris}}]{smith2015}
{Smith}, R.~L., {Pontoppidan}, K.~M., {Young}, E.~D., \& {Morris}, M.~R. 2015,
  \apj, 813, 120, \dodoi{10.1088/0004-637X/813/2/120}

\bibitem[{{Sorahana} \& {Yamamura}(2012)}]{sorahana2012}
{Sorahana}, S., \& {Yamamura}, I. 2012, \apj, 760, 151,
  \dodoi{10.1088/0004-637X/760/2/151}

\bibitem[{{Tennyson} {et~al.}(2016){Tennyson}, {Yurchenko}, {Al-Refaie},
  {Barton}, {Chubb}, {Coles}, {Diamantopoulou}, {Gorman}, {Hill}, {Lam},
  {Lodi}, {McKemmish}, {Na}, {Owens}, {Polyansky}, {Rivlin}, {Sousa-Silva},
  {Underwood}, {Yachmenev}, \& {Zak}}]{tennyson2016}
{Tennyson}, J., {Yurchenko}, S.~N., {Al-Refaie}, A.~F., {et~al.} 2016, Journal
  of Molecular Spectroscopy, 327, 73, \dodoi{10.1016/j.jms.2016.05.002}

\bibitem[{{The LUVOIR Team}(2019)}]{luvoir2019}
{The LUVOIR Team}. 2019, arXiv e-prints, arXiv:1912.06219,
  \dodoi{10.48550/arXiv.1912.06219}

\bibitem[{{Todorov} {et~al.}(2016){Todorov}, {Line}, {Pineda}, {Meyer},
  {Quanz}, {Hinkley}, \& {Fortney}}]{todorov2016}
{Todorov}, K.~O., {Line}, M.~R., {Pineda}, J.~E., {et~al.} 2016, \apj, 823, 14,
  \dodoi{10.3847/0004-637X/823/1/14}

\bibitem[{Vehtari {et~al.}(2017)Vehtari, Gelman, \& Gabry}]{vehtari2017}
Vehtari, A., Gelman, A., \& Gabry, J. 2017, Statistics and computing, 27, 1413

\bibitem[{{Visser} {et~al.}(2009){Visser}, {van Dishoeck}, \&
  {Black}}]{visser2009}
{Visser}, R., {van Dishoeck}, E.~F., \& {Black}, J.~H. 2009, \aap, 503, 323,
  \dodoi{10.1051/0004-6361/200912129}

\bibitem[{{Voronin} {et~al.}(2010){Voronin}, {Tennyson}, {Tolchenov},
  {Lugovskoy}, \& {Yurchenko}}]{voronin2010}
{Voronin}, B.~A., {Tennyson}, J., {Tolchenov}, R.~N., {Lugovskoy}, A.~A., \&
  {Yurchenko}, S.~N. 2010, \mnras, 402, 492,
  \dodoi{10.1111/j.1365-2966.2009.15904.x}

\bibitem[{{Welbanks} {et~al.}(2023){Welbanks}, {McGill}, {Line}, \&
  {Madhusudhan}}]{welbanks2023}
{Welbanks}, L., {McGill}, P., {Line}, M., \& {Madhusudhan}, N. 2023, \aj, 165,
  112, \dodoi{10.3847/1538-3881/acab67}

\bibitem[{{Wilson}(1999)}]{wilson1999}
{Wilson}, T.~L. 1999, Reports on Progress in Physics, 62, 143,
  \dodoi{10.1088/0034-4885/62/2/002}

\bibitem[{{Wouterloot} {et~al.}(2008){Wouterloot}, {Henkel}, {Brand}, \&
  {Davis}}]{wouterloot2008}
{Wouterloot}, J.~G.~A., {Henkel}, C., {Brand}, J., \& {Davis}, G.~R. 2008,
  \aap, 487, 237, \dodoi{10.1051/0004-6361:20078156}

\bibitem[{{Zhang} {et~al.}(2022){Zhang}, {Snellen}, {Brogi}, \&
  {Birkby}}]{zhang2022}
{Zhang}, Y., {Snellen}, I. A.~G., {Brogi}, M., \& {Birkby}, J.~L. 2022,
  Research Notes of the American Astronomical Society, 6, 194,
  \dodoi{10.3847/2515-5172/ac9309}

\bibitem[{{Zhang} {et~al.}(2021{\natexlab{a}}){Zhang}, {Snellen}, \&
  {Molli{\`e}re}}]{zhang2021_bd}
{Zhang}, Y., {Snellen}, I. A.~G., \& {Molli{\`e}re}, P. 2021{\natexlab{a}},
  \aap, 656, A76, \dodoi{10.1051/0004-6361/202141502}

\bibitem[{{Zhang} {et~al.}(2021{\natexlab{b}}){Zhang}, {Snellen}, {Bohn},
  {Molli{\`e}re}, {Ginski}, {Hoeijmakers}, {Kenworthy}, {Mamajek}, {Meshkat},
  {Reggiani}, \& {Snik}}]{zhang2021_nature}
{Zhang}, Y., {Snellen}, I. A.~G., {Bohn}, A.~J., {et~al.} 2021{\natexlab{b}},
  \nat, 595, 370, \dodoi{10.1038/s41586-021-03616-x}

\bibitem[{{Zhou} {et~al.}(2022){Zhou}, {Bowler}, {Apai}, {Kataria}, {Morley},
  {Bryan}, {Skemer}, \& {Benneke}}]{zhou2022}
{Zhou}, Y., {Bowler}, B.~P., {Apai}, D., {et~al.} 2022, \aj, 164, 239,
  \dodoi{10.3847/1538-3881/ac9905}

\bibitem[{{Zhou} {et~al.}(2020){Zhou}, {Bowler}, {Morley}, {Apai}, {Kataria},
  {Bryan}, \& {Benneke}}]{zhou2020}
{Zhou}, Y., {Bowler}, B.~P., {Morley}, C.~V., {et~al.} 2020, \aj, 160, 77,
  \dodoi{10.3847/1538-3881/ab9e04}

\end{thebibliography}
\bibliographystyle{aasjournal}



\end{document}